\documentclass[aps.prd,twocolumn,floatfix,superscriptaddress,showpacs]{revtex4-1}

\usepackage{graphicx}% Include figure files
\usepackage{bm}% bold math
\usepackage{bbold}
\usepackage{amssymb,amsfonts,amsmath}
\usepackage{hyperref}% add hypertext capabilities
\usepackage{color}
\usepackage{amsfonts}
\usepackage{subfigure}
\usepackage{array}
\usepackage{ulem}

\newcolumntype{L}{>{\centering\arraybackslash}m{1.8cm}}
\newcolumntype{S}{>{\centering\arraybackslash}m{1.cm}}
\newcolumntype{M}{>{\centering\arraybackslash}m{1.5cm}}

\def\Fig#1{Fig.~\ref{#1}}

\def\Eq#1{Eq.~(\ref{#1})}
\def\eq#1{(\ref{#1})}
\def\eqref#1{(\ref{#1})}

\def\tab#1{Table~\ref{#1}}

\def\sec#1{Sec.~\ref{#1}}

\def\lA0{{\langle A_0 \rangle}}
\def\bA0{{\bar{A}_0}}

\def\0#1#2{\frac{#1}{#2}}

\graphicspath{{./figures/}{./}}

\begin{document}

\preprint{}

\title{Identifying the QCD Phase Transitions via the Gravitational Wave Frequency}
%
% Force line breaks with \\

\author{Wei-jie Fu}
%\email{wjfu@dlut.edu.cn}
\affiliation{Institute of Theoretical Physics, School of Physics \&
  Optoelectronic Technology, Dalian University of Technology, Dalian, 116024,
  P.R. China}

\author{Zhan Bai}
%\email[]{}
\affiliation{Department of Physics and State Key Laboratory of Nuclear Physics and Technology, Peking University, Beijing 100871, China}

\author{Yu-xin Liu}
\email[]{yxliu@pku.edu.cn}
\affiliation{Department of Physics and State Key Laboratory of Nuclear Physics and Technology, Peking University, Beijing 100871, China}
\affiliation{Collaborative Innovation Center of Quantum Matter, Beijing 100871, China}
\affiliation{Center for High Energy Physics, Peking University, Beijing 100871, China}


\begin{abstract}
We investigate the nonradial oscillations of newly born neutron stars (NSs) and strange quark stars (SQSs). This is done with the relativistic nuclear field theory with hyperon degrees of freedom employed to describe the equation of state for the stellar matter in NSs, and with both the MIT bag model and the Nambu--Jona-Lasinio model adopted to construct the configurations of the SQSs. We find that the gravitational-mode ($g$-mode) eigenfrequencies of newly born SQSs are about one order of magnitude lower than those of NSs, which is independent of models implemented to describe the equation of state for the strange quark matter. Meanwhile the eigenfrequencies of the other modes of nonradial oscillations, {\it e.g.}, fundamental ($f$)- and pressure ($p$)-modes, are much larger than those of the $g$-mode. In the light of the first direct observation of gravitational waves, it is promising to employ the gravitational waves to identify the QCD phase transition in high density strong interaction matter.
\end{abstract}

                             % Classification Scheme.

\maketitle

%\tableofcontents

\section{\label{sec:Intro}Introduction}

The first direct observation of gravitational waves from a binary black hole (BH) merger on September 14, 2015 by LIGO Scientific Collaboration and Virgo Collaboration~\cite{Abbott:2016blz}, has opened up a new era for the astronomy, viz. the gravitational wave astronomy. The potential high-frequency sources sought by earth-based gravitational-wave detectors, such as LIGO, VIRGO, GEO600, TAMA300, include not only the BH/BH merger and ringdown, as the source of the gravitational-wave event GW150914~\cite{Abbott:2016blz}, but also the tidal disruption of a neutron star (NS) by its companion BH in NS/BH binaries, spinning NSs, type-II supernovae, proto-neutron-stars produced by the accretion-induced collapse of white dwarf stars, etc., for more details about the gravitational-wave sources, see, {\it e.g.}, Refs.~\cite{Cutler:2002me,Andersson:2009yt}. Remarkably, NSs, besides the BHs, are important sources of high-frequency gravitational waves.

On the other hand, NSs, as compact objects in the cosmos, are ideal laboratories for physics in extreme environments, such as the strong gravitational fields, QCD at high density, high magnetic fields, etc., see, {\it e.g.}, Ref.~\cite{Glendenning:2000} for details. In fact, there have been a longstanding debate whether the QCD phase transition takes place in the interior of these compact stars, {\it i.e.}, whether strange quark stars (SQSs) are also potential candidates for them~\cite{Itoh:1970uw,Freedman:1977gz,Witten:1984rs,Alcock:1986hz,Weber:2004kj,Alford:2006vz,Lattimer:2006xb}.
To answer this question, one can, of course, resort to earth-based heavy-ion collision experiments, such as at the Relativistic Heavy-Ion Collider (RHIC)~\cite{Adams:2005dq,Adcox:2004mh}, the Large Hadron Collider (LHC) at CERN~\cite{Aamodt:2010pa},   the Facility for Antiproton and Ion Research (FAIR) at GSI and the Nuclotron-based Ion Collider fAcility (NICA) at Dubna~\cite{Batyuk:2016}.
Indeed significant progresses have been made in recent years from experimental measurements, for instance, the Beam Energy Scan (BES) program at RHIC~\cite{Adamczyk:2013dal,Adamczyk:2014fia,Luo:2015ewa}, aiming to get hold of the existence and location of the critical end point (CEP) in the QCD phase diagram~\cite{Luo:2015ewa,Stephanov:2007fk}. The QCD phase structure, however, is far from being unveiled, because of the notorious sign problem at finite chemical potential, which prevents the first-principle lattice QCD simulations from getting access to high density regime~\cite{Aarts:2015tyj}. In turn, the gravitational wave astronomy provides a potential, new approach to study the QCD at extreme densities.

Many observable differences have been proposed to identify the QCD phase transition in compact stars, {\it i.e.}, to distinguish SQSs from NSs. SQSs are found to have larger dissipation rate of radial vibrations~\cite{Wang:1985tg} and higher bulk viscosity~\cite{Haensel:1989}.
The cooling of SQSs is more rapid than that of NSs within the first 30 years~\cite{Schaab:1997hx}. It was found that the spin rate of SQSs can be much closer to the Kepler limit than that of NSs~\cite{Madsen:1992sx}.
In Ref.~\cite{Bauswein:2009im} three-dimensional relativistic hydrodynamical simulations of the coalescence of SQSs are performed, and relevant gravitational-wave signals are extracted in comparison to NS coalescence. Recently, it has been found that the observation of old SQSs can set important limits on the scattering cross sections between the light quarks and the bosonic non-interacting dark matter~\cite{Zheng:2016ygg}.

We have studied the eigenfrequencies of the gravitational-mode, {\it i.e.}, the $g$-mode, oscillation of newly born SQSs and NSs in Ref.~\cite{Fu:2008bu}, and found that the eigenfrequencies of SQSs are much lower than those of NSs, because the components of a SQS are almost extremely relativistic particles while nucleons in a NS are non-relativistic. Furthermore, simulations of core-collapse supernovae have indicated that $g$-mode oscillations of the supernovae core may be excited~\cite{Burrows:2005dv,Burrows:2006ci}, and serve as efficient sources of gravitational waves~\cite{Ott:2006qp,Misner:1970,Owen:2005fn,Lai:2006pr}, for more related discussions about the simulations of core-collapse supernovae, see {\it e.g.}~\cite{Blondin:2002sm,Foglizzo:2001ke}.
In this work we extend our studies in Ref.~\cite{Fu:2008bu}  to not only considering the hyperon degrees of freedom for the equation of state (EOS) of the matter in NSs
but also including other nonradial modes, such as the $f$- and $p$-modes, for comparison.
Details about the calculations will be presented, and the dependence of the eigenfrequencies on different equations of state will also be discussed.

The paper is organized as follows. In \sec{sec:nonradial} the nonradial oscillations of a nonrotating, unmagnetized, and fluid star are briefly described. In \sec{sec:NSs} the model of newly born NSs is constructed and discussed. In \sec{sec:MIT} we employ the MIT bag model to construct the configurations of the newly born SQSs. In \sec{sec:numerical} we present the numerical results and discussions. In \sec{sec:NJL} the Nambu--Jona-Lasinio model is adopted to described the EOS of state of the strange quark matter, and relevant calculated results of the $g$-mode eigenfrequencies are presented. Section~\ref{sec:sum} summarizes our conclusions.

\section{\label{sec:nonradial}Nonradial Oscillations of a Star}

In this section, we reiterate briefly the nonradial oscillations of a nonrotating, unmagnetized, and fluid star (for details, see {\it e.g.}, Refs.~\cite{Reisenegger:1992,Cox:1980,Lai:1993di}).
Assuming the equilibrium configuration of the star to be spherically symmetric, the pressure $p_{0}^{}$, density $\rho_{0}^{}$ and the gravitational potential $\phi_{0}^{}$ are functions of only the radial coordinate $r$. A vector field of displacement $\bm{\xi}(\bm{r},t)$ describes the oscillations, and the Eulerian perturbations of the pressure, density and the potential are $\delta p$, $\delta \rho$ and $\delta \phi$, respectively. The Newtonian equation of motion (EoM) reads
\begin{align}
-\nabla p-\rho\nabla\phi=\rho\frac{\partial^{2}\bm{\xi}}{\partial
t^{2}} \, ,\label{eq:eom1}
\end{align}
where $p=p_{0}^{}+\delta p$, as well as $\rho$ and $\phi$. Assuming the perturbations are small, and only keeping linear terms of the perturbations, we are led to
\begin{align}
-\nabla \delta p - \delta \rho\nabla\phi_{0}^{} - \rho_{0}^{}\nabla \delta
\phi = \rho_{0}^{} \frac{\partial^{2}\bm{\xi}}{\partial t^{2}} \, ,      \label{eq:eom2}
\end{align}
where we have employed equilibrium equation, {\it i.e.},
\begin{align}
-\nabla p_{0}-\rho_{0}\nabla\phi_{0}=0 \, .   \label{eq:equi}
\end{align}
In a slight abuse of notation we employ $p$, $\rho$ and $\phi$ to denote the equilibrium values from now on, rather than those with subscript $_0$.

Furthermore, we have continuity equation which reads
\begin{align}
\delta \rho+\nabla\cdot(\rho\bm{\xi}) = 0 \, . \label{eq:continuity}
\end{align}
Here, it is more convenient to adopt the Lagrangian perturbations $\Delta$, which is related with the Eulerian formalism through the following relation, {\it i.e.},
\begin{align}
\Delta=\delta+\bm{\xi}\cdot\nabla\,.
\end{align}
Then Eq.~(\ref{eq:continuity}) follows readily as
\begin{align}
\Delta \rho+\rho\nabla\cdot\bm{\xi}=0 \, . \label{eq:continuity2}
\end{align}

For a stellar oscillation with an eigenfrequency $\omega$, all the perturbative quantities, {\it i.e.}, $\bm{\xi}$, $\delta p$, $\delta\rho$, and $\delta \phi$, are dependent on time through a temporal factor $e^{-i\omega t}$. Inserting this factor into the EoM (in Eq.~(\ref{eq:eom2})), we arrive at
\begin{align}
-\omega^{2}\bm{\xi}=-\frac{\nabla(\delta p)}{\rho}-\frac{\delta
\rho}{\rho}\nabla\phi-\nabla(\delta\phi)\, .\label{eq:eom3}
\end{align}
Furthermore, we have the linearized Poisson equation for the gravitational potential, {\it i.e.},
\begin{align}
\nabla^{2}\delta\phi=4\pi G \delta\rho \, ,  \label{eq:Poisson}
\end{align}
where $G$ is the gravitational constant. Eqs.~(\ref{eq:continuity2})--(\ref{eq:Poisson}) constitute a closed set of equations of nonradial oscillations for a nonrotating, fluid star. In the following we will rewrite those vectorial equations in component formalism. The transversal component of displacement ${\bm{\xi}}_{\bot}$, {\it i.e.}, perpendicular to the radial direction, is easily deduced from Eq.~(\ref{eq:eom3}), which reads
\begin{align}
{\bm{\xi}}_{\bot}=\frac{\nabla_{\bot}(\delta
p)}{\omega^{2}\rho}+\frac{\nabla_{\bot}(\delta\phi)}{\omega^{2}}\,,\label{eq:xiPer}
\end{align}
with
\begin{align}
\nabla_{\bot}=\frac{1}{r}\frac{\partial}{\partial\theta}\bm{e}_{\theta}
+\frac{1}{r\sin\theta}\frac{\partial}{\partial\varphi}\bm{e}_{\varphi}\,.
\end{align}
Substituting Eq.~(\ref{eq:xiPer}) into Eq.~(\ref{eq:continuity2}), one arrives at
\begin{align}
\frac{\Delta \rho}{\rho}+\frac{1}{r^{2}}\frac{\partial}{\partial
r}(r^{2}\xi_{r})+\frac{1}{\omega^{2}\rho}\nabla_{\bot}^{2}(\delta
p)+\frac{1}{\omega^{2}}\nabla_{\bot}^{2}(\delta \phi)=0  \,  ,\label{eq:continuity3}
\end{align}
with
\begin{align}
\nabla_{\bot}^{2}=\frac{1}{r^{2}\sin\theta}\frac{\partial}{\partial\theta}\left(\sin\theta\frac{\partial}{\partial\theta}\right)
+\frac{1}{r^{2}\sin^{2}\theta}\frac{\partial^{2}}{\partial\varphi^{2}}\, .
\end{align}

Both the $\delta p$ and $\delta\phi$  in an eigenmode of the oscillations can be factorized into products of a spherical harmonic $Y_{lm}(\theta,\varphi)$  and a radial coordinate dependent function. Thus, for an eigenmode, Eq.~(\ref{eq:continuity3}) is simplified to the following equation:
\begin{align}
\frac{1}{r^{2}}\frac{\partial}{\partial
r}(r^{2}\xi_{r})-\frac{l(l+1)}{\omega^{2}r^{2}}\frac{\delta
p}{\rho}-\frac{l(l+1)}{\omega^{2}r^{2}}\delta \phi+\frac{\Delta
\rho}{\rho}=0\, .\label{eq:continuity4}
\end{align}

Considering the adiabatic sound speed $c_{s}$ defined as
\begin{align}
c_{s}^{2}\equiv\left(\frac{\partial p}{\partial
\rho}\right)_{\mathrm{adia}}=\frac{\Delta p}{\Delta \rho} \, , \label{eq:cs}
\end{align}
one has
\begin{align}
\Delta \rho & = \frac{\Delta
p}{c_{s}^{2}}=\frac{(\delta+\bm{\xi}\cdot\nabla)p}{c_{s}^{2}}\nonumber\\
& = \frac{\delta p-\xi_{r}g\rho}{c_{s}^{2}}\,,\label{eq:Delrho}
\end{align}
where we have employed the local gravitational acceleration $\bm{g}\equiv-\nabla\phi$.
Inserting Eq.~(\ref{eq:Delrho}) into Eq.~(\ref{eq:continuity4}), we are left with
\begin{align}
\frac{\partial}{\partial r}(r^{2}\xi_{r}) =&
\, \frac{g}{c_{s}^{2}}(r^{2}\xi_{r})
+\left[\frac{l(l+1)}{\omega^{2}}-\frac{r^{2}}{c_{s}^{2}}\right]
\left(\frac{\delta
p}{\rho}\right)\nonumber\\ &+\frac{l(l+1)}{\omega^{2}}\delta\phi\,.
\label{eq:Oscil1}
\end{align}

The radial component of the EoM in Eq.~(\ref{eq:eom3}) is readily obtained, which is given by
\begin{align}
-\omega^{2}\xi_{r}=-\frac{1}{\rho}\frac{\partial(\delta p)}{\partial
r}-\frac{\delta \rho}{\rho}g-\frac{\partial(\delta\phi)}{\partial
r} \, .\label{eq:radial}
\end{align}

For the $\delta \rho$, one has
\begin{align}
\delta \rho& = (\Delta-\bm{\xi}\cdot\nabla)\rho=\frac{\Delta
p}{c_{s}^{2}}-\xi_{r}\frac{d \rho}{d r}\nonumber\\
& = \frac{\Delta p}{c_{s}^{2}}-\frac{\xi_{r}}{c_{e}^{2}}\frac{d
p}{d r}\nonumber\\
& = \left(\frac{\rho}{c_{s}^{2}}\right)\left(\frac{\delta
p}{\rho}\right)+g^{2}\left(\frac{1}{c_{e}^{2}}-\frac{1}{c_{s}^{2}}\right)\left(\frac{\rho}{g
r^{2}}\right)(r^{2}\xi_{r})  \, ,  \label{eq:drho}
\end{align}
where we have introduced another sound speed, {\it i.e.}, equilibrium sound speed $c_{e}$, defined by
\begin{align}
c_{e}^{2}\equiv\left(\frac{d p}{d
\rho}\right)_{\mathrm{equi}}=\frac{d p/d r}{d \rho/d r}  \,  .   \label{eq:ce}
\end{align}
In the same time, we have
\begin{align}
\frac{\partial(\delta p)}{\partial r}
= \rho\frac{\partial}{\partial r}\left(\frac{\delta p}{\rho}\right) -
\frac{g\rho}{c_{e}^{2}}\left(\frac{\delta p}{\rho}\right)    \,    .   \label{eq:ddelp}
\end{align}
Substituting Eqs.~(\ref{eq:drho}) and (\ref{eq:ddelp}) into Eq.~(\ref{eq:radial}), one is led to
\begin{align}
\frac{\partial}{\partial r}\left(\frac{\delta p}{\rho}\right) =
\frac{\omega^{2}-\omega_{BV}^{2}}{r^{2}}(r^{2}\xi_{r})
+\frac{\omega_{BV}^{2}}{g}\left(\frac{\delta
p}{\rho}\right)-\frac{\partial \delta\phi}{\partial r}   \,   ,
\label{eq:Oscil2}
\end{align}
where $\omega_{BV}^{}$ is the Brunt-V\"{a}is\"{a}l\"{a} frequency, which is given by
\begin{align}
\omega_{BV}^{2}=g^{2}\left(\frac{1}{c_{e}^{2}}-\frac{1}{c_{s}^{2}}\right)
\, .\label{eq:BV}
\end{align}

In the same way, the Poisson equation for perturbations of the gravitational potential in Eq.~(\ref{eq:Poisson}) can be reformulated as
\begin{align}
\frac{1}{r^{2}}\frac{\partial}{\partial r}\left(r^{2}\frac{\partial
\delta\phi}{\partial r}\right) = & \, 4\pi
G\left[\omega_{BV}^{2}\left(\frac{\rho}{g
r^{2}}\right)(r^{2}\xi_{r})\right.\nonumber\\
& \left.+\left(\frac{\rho}{c_{s}^{2}}\right)\left(\frac{\delta
p}{\rho}\right)\right]+\frac{l(l+1)\delta\phi}{r^{2}} \,
.\label{eq:Poisson2}
\end{align}
In the following calculations, we adopt the so called Cowling approximation~\cite{Cowling:1941}, {\it i.e.}, neglecting the perturbations of the gravitational potential $\delta\phi$.

For the convenience of numerical computations in the following, we parameterize the displacement vector as:
\begin{align}
\bm{\xi}=(\eta_{r}\bm{e}_{r}+r\eta_{\bot}\nabla_{\bot})Y_{lm}(\theta,\varphi)e^{-i\omega
t}   \,  .  \label{eq:eta}
\end{align}
Recalling Eq.~(\ref{eq:xiPer}), we have
\begin{align}
\xi_{r}(r,\theta,\varphi,t) & = \eta_{r}(r)Y_{lm}(\theta,\varphi)e^{-i\omega
t},\nonumber\\
\frac{\delta p(r,\theta,\varphi,t)}{r\omega^{2}\rho}& =
\eta_{\bot}(r)Y_{lm}(\theta,\varphi)e^{-i\omega t}   \,  .   \label{eq:displa}
\end{align}
Inserting these two equations into Eqs.~(\ref{eq:Oscil1}) and (\ref{eq:Oscil2}), we obtain
\begin{align}
\frac{d \eta_{r}}{d r}\! & = \left(\frac{g r}{c_{s}^{2}}-2\right)\frac{\eta_{r}^{}}{r}
+\left[l(l+1)-\frac{\omega^{2}r^{2}}{c_{s}^{2}}\right]\frac{\eta_{\bot}^{}}{r}  \,  , \label{eq:Osc1}
\\[1mm]
\frac{d \eta_{\bot}}{d r}\! & =
\left(1-\frac{\omega_{BV}^{2}}{\omega^{2}}\right)\frac{\eta_{r}^{}}{r}
+\left(\frac{\omega_{BV}^{2}r}{g}-1\right)\frac{\eta_{\bot}^{}}{r} \, .
\label{eq:Osc2}
\end{align}
In order to determine the eigenfrequency of a mode of oscillations, one also needs the boundary conditions which, in the stellar center, is given by
\begin{align}
\eta_{r}^{} = A\,lr^{l-1}, \qquad \eta_{\bot}^{} = A\,r^{l-1} \qquad (r\rightarrow
0)\,,\label{eq:bound1}
\end{align}
where $A$ is an arbitrary constant, and is related with the amplitude of the oscillation. While at the outer surface of the star, the Lagrangian perturbation of the pressure is vanishing, {\it i.e.},
\begin{align}
\Delta p & =(\delta +\bm{\xi}\cdot\nabla)p\nonumber\\
&=(\omega^{2}r\eta_{\bot}^{} - g\eta_{r}^{})\rho=0  \qquad
(r=R)\,.\label{eq:bound2}
\end{align}

In our calculations we employ the Newtonian hydrostatic equations, rather than those of general relativity, to determine the equilibrium configurations of newly born NSs and SQSs, in order to coincide with Eqs.~(\ref{eq:Oscil1}) and (\ref{eq:Oscil2}) or Eqs.~(\ref{eq:Osc1}) and (\ref{eq:Osc2}). Then, the adiabatic, equilibrium sound speed, $c_{s}$ and $c_{e}$ respectively, and the Brunt-V\"{a}is\"{a}l\"{a} frequency $\omega_{BV}^{}$ can be obtained as functions of the radial coordinate $r$.

Solving the equations of nonradial oscillations with Cowling approximation in Eqs.~(\ref{eq:Osc1}) and (\ref{eq:Osc2}), together with the boundary conditions in Eqs.~(\ref{eq:bound1}) and (\ref{eq:bound2}), one obtains three classes of modes of oscillations, which are the gravitational-mode ($g$-mode), fundamental-mode ($f$-mode), and the pressure-mode ($p$-mode), respectively. Generally speaking, stellar oscillations of $g$-mode originate from the buoyancy in the star and, thus, the eigenfrequency of $g$-mode is intimately linked to Brunt-V\"{a}is\"{a}l\"{a} frequency $\omega_{BV}^{}$. $f$-mode and $p$-mode, however, result from the pressure inside the star, and $f$-mode is in fact a particular mode of $p$-modes, with the number of the radial node being zero. The eigenfrequencies of $f$-mode and $p$-mode are therefore related only with the sound speed of the stellar matter. Of the three classes of oscillation modes, $g$-mode is our main focus in the following, because it may be a source of the gravitational wave~\cite{Burrows:2005dv,Burrows:2006ci,Ott:2006qp,Misner:1970,Owen:2005fn,Lai:2006pr}.

\section{\label{sec:NSs}Modelling the Newly Born Neutron Stars}

In this section as well as those below, we will construct models of newly born NSs and SQSs, respectively. In comparison with ordinary NSs after cooling, such as pulsars, newly born NSs in a core-collapse supernova feature two significant properties: one is the high temperature. The inner temperature of pulsars is of order of $10^9\,\mathrm{K}$ and below, while that of newly born NSs can be higher than $10^{11}\,\mathrm{K}$. The other is the high abundance of neutrinos or leptons. In the first tens of seconds after the core bounce in a core-collapse supernova, neutrinos and antineutrinos of different flavors are trapped in the stellar matter of high temperatures and densities, since their mean free paths are smaller than the size of NSs~\cite{Bethe:1990mw}. Here we will not delve into the evolution and structure of the newly born NSs, for more details, see, {\it e.g.}, Refs.~\cite{Burrows:1986me,Pons:1998mm,Keil:1995hw,Prakash:1996xs,Pons:2000xf}. In this work we extend our former calculations in Ref.~\cite{Fu:2008bu} to allow for the existence of hyperons, besides nucleons. A fluid element inside the star can be described by three independent variables, such as the baryon density $\rho_{B}^{}$, the entropy per baryon $S$, and the lepton fraction $Y_{L}=Y_{e}+Y_{\nu_{e}}$ ($Y_{i}={\rho_{i}^{}}/{\rho_{B}^{}}$). Surely, one can use the electron fraction $Y_{e}$ in lieu of $Y_{L}$, which are related with each other through beta equilibrium conditions. In our calculations we employ the relativistic mean field (RMF) theory at finite temperature to describe the interactions among baryons~\cite{Serot:1984ey,Glendenning:2000}, while assume all other components of the star to be non-interacting. The Lagrangian density of the RMF is given by
%\begin{widetext}
\begin{align}
\mathcal{L}  = &
\sum_i\bar{\Psi}_{i}\big[i\gamma_{\mu}\partial^{\mu}\negmedspace-\negmedspace
 \left(m_{i}\negmedspace -\negmedspace g_{\sigma i}\sigma\,\right)\negmedspace -\negmedspace g_{\omega i} \gamma_{\mu} \omega^{\mu}\negmedspace -\negmedspace {g_{\rho i}^{}} \gamma^{\mu}\bm{t}_i\cdot {\bm{\rho}_{\mu}^{}} \big]\Psi_{i}\nonumber\\
& + \frac{1}{2}\left(\partial_{\mu}\sigma\partial^{\mu}\sigma
 -m_{\sigma}^{2}\sigma^{2}\right) \nonumber \\
 & -\frac{1}{3}b\,m_{N}^{}({g_{\sigma N}^{}} \sigma)^{3}
 -\frac{1}{4}c({g_{\sigma N}^{}} \sigma)^{4}\nonumber\\
& - \frac{1}{4}\omega_{\mu\nu}\omega^{\mu\nu}+\frac{1}{2}m^{2}_{\omega}
 \omega_{\mu}\omega^{\mu} \nonumber \\
 & -\frac{1}{4} {\bm{\rho}_{\mu\nu}^{}} \cdot\bm{\rho}^{\mu\nu}
 +\frac{1}{2}m^{2}_{\rho} {\bm{\rho}_{\mu}^{}} \cdot
 \bm{\rho}^{\mu}
,\label{eq:RMFlagrangian}
\end{align}
where $\Psi_{i}$ ($i=p,n,\Lambda,\Sigma^{\pm,0},\Xi^{-,0}$) are the baryon octet fields, which interact through exchanging mesons with coupling strength $g_{ji}^{}$ ($j$ refers to the meson $\sigma$, $\rho$ and $\omega$, $i$ stands for the baryons). The isoscalar-scalar meson $\sigma$ provides attractive interactions between baryons, while the isoscalar-vector meson $\omega$ accounts for repulsive interactions. The isovector-vector $\rho$ meson distinguishes between baryons with different isospin, and thus plays a significant role in determining, {\it e.g.}, the symmetry energy of the nuclear matter.
Note that self-interactions of $\sigma$-meson in Eq.~(\ref{eq:RMFlagrangian}) are also imperative to the correct description of the nuclear matter in the RMF formalism. $\omega_{\mu\nu}^{}$ and $\bm{\rho}_{\mu\nu}^{}$ in Eq.~(\ref{eq:RMFlagrangian}) are the vector meson field tensors, which read
\begin{align}
 \omega_{\mu\nu} & \equiv
\partial_{\mu}\omega_{\nu}-\partial_{\nu}\omega_{\mu},\\
 \bm{\rho}_{\mu\nu} & \equiv
\partial_{\mu}\bm{\rho}_{\nu}-\partial_{\nu}\bm{\rho}_{\mu},
\end{align}
and $t^a$ is the Pauli matrices with $\mathrm{tr}(t^at^b)=\delta_{ab}/2$.
In the mean field approximation, the Dirac equation for baryon $i$ follows readily as
\begin{align}
\left[i\gamma_{\mu}\partial^{\mu} - \left(m_{i}^{} - {g_{\sigma i}^{}} \sigma   \right)
- {g_{\omega i}^{}} \gamma^{0} {\omega_{0}^{}}
   - {g_{\rho i}^{}} \gamma^{0} t_{3i}^{} {\rho_{03}^{}} \right]\Psi_{i}=0,
   \label{eq:Dirac}
\end{align}
and thus, the dispersion relation for baryons reads
\begin{align}
E_{i}&=\sqrt{p^{2}+{m^{*}_{i}}^{2}} + {g_{\omega i}^{}} \omega_{0}+{g_{\rho i}^{}} {\rho_{03}^{}}t_{3i} \,,
%\nonumber\\
%\sout{E_{n}&=\sqrt{p^{2}+{m^{*}_{n}}^{2}} + {g_{\omega}^{}} \omega_{0} - \frac{1}{2} {g_{\rho}^{}} {\rho_{03}^{}}} \,,
\label{eq:dispersion}
\end{align}
with effective mass
\begin{align}
m^{*}_{i} = m_{i} - {g_{\sigma i}^{}} \sigma\,. \label{eq:mass}
\end{align}
and $t_{3i}^{}$ is the 3-component of the isospin of baryon $i$.
%
%, {\it e.g.} ${t_{3}}_i=\pm 1/2$ for $i=p,n$.
%

The thermodynamical potential density for baryons in the formalism of RMF is given by
\begin{align}
\Omega_{B}&=-T\sum_{i}2\int\frac{d^{3}\bm{k}}{(2\pi)^{3}}\bigg\{
\ln\Big[1+e^{-(E^{*}_{i}(k)-\mu^{*}_{i})/T}\Big]\nonumber\\
& +\ln\Big[1+e^{-(E^{*}_{i}(k)+\mu^{*}_{i})/T}\Big]\bigg\} \nonumber \\ & +\frac{1}{2}m_{\sigma}^{2}\sigma^{2} + \frac{1}{3}b\,m_{N}^{} ( {g_{\sigma N}^{}} \sigma)^{3}
\nonumber\\
& +\frac{1}{4}c( {g_{\sigma N}^{}} \sigma)^{4} - \frac{1}{2}m^{2}_{\omega}
 \omega_{0}^{2} - \frac{1}{2}m^{2}_{\rho}
 \rho_{03}^{2}\,,
\label{eq:thermo}
\end{align}
with temperature $T$ and $E^{*}_{i}=(p^{2}+{m^{*}_{i}}^{2})^{1/2}$. $\mu^{*}_{i}$ is the effective chemical potential, which is defined as
\begin{align}
\mu^{*}_{i}=\mu_{i} - {g_{\omega i}^{}} \omega_{0}-{t_{3}}_{i} {g_{\rho i}^{}} {\rho_{03}^{}} \,.\label{eq:chemi}
\end{align}
Employing Eq.~(\ref{eq:thermo}), one can easily obtain other thermodynamical quantities, the number density, for instance, reading
\begin{align}
\rho_{i}^{} &=-\frac{\partial \Omega_{B}}{\partial {\mu_{i}^{}} }\nonumber\\
&=2\int\frac{d^{3}\bm{k}}{(2\pi)^{3}} \Big{[} f(E^{*}_{i}(k))-\bar{f}(E^{*}_{i}(k)) \Big{]} \,,\label{eq:density}
\end{align}
with the Fermi-Dirac distribution functions being
\begin{align}
f(E^{*}_{i}(k))&=\frac{1}{\exp[(E^{*}_{i}(k)-\mu^{*}_{i})/T]+1}\,,\label{eq:fermi}\\
\bar{f}(E^{*}_{i}(k))&=\frac{1}{\exp[(E^{*}_{i}(k)+\mu^{*}_{i})/T]+1} \, . \label{eq:antifermi}
\end{align}
And the entropy density is given by
\begin{align}
S_{B}&=-\frac{\partial \Omega_{B}}{\partial T}\nonumber\\
&=\frac{2}{T}\! \int \! \! \frac{d^{3}\bm{k}}{(2\pi)^{3}} \!
\sum_{i} \! \Big{[} \Big{(} E^{*}_{i}(k)+\frac{k}{3}\frac{d E^{*}_{i}(k)}{dk}-\mu^{*}_{i}\Big{)} f(E^{*}_{i}(k))\nonumber\\
&\quad\;+\Big{(} E^{*}_{i}(k)+\frac{k}{3}\frac{d E^{*}_{i}(k)}{d
k}+\mu^{*}_{i}\Big{)} \bar{f}(E^{*}_{i}(k)) \Big{]} \, .
\label{eq:entropy}
\end{align}
The pressure is given by
\begin{align}
p_{B}^{} =&-\Omega_{B}\nonumber\\
=&\sum_{i}2\int\frac{d^{3}\bm{k}}{(2\pi)^{3}} \frac{k}{3}\frac{d
E^{*}_{i}(k)}{dk} \Big{[} f(E^{*}_{i}(k))+\bar{f}(E^{*}_{i}(k)) \Big{]} \nonumber\\
&-\frac{1}{2}m_{\sigma}^{2}\sigma^{2}-\frac{1}{3}b\,m_{N}( {g_{\sigma N}^{}} \sigma)^{3}
 -\frac{1}{4}c({g_{\sigma N}^{}} \sigma)^{4}\nonumber\\
&+\frac{1}{2}m^{2}_{\omega}
 \omega_{0}^{2}+\frac{1}{2}m^{2}_{\rho}
 \rho_{03}^{2}\,. \label{eq:pressure}
\end{align}
Moreover, the energy density for the strongly interacting baryons takes the form:
\begin{align}
\varepsilon_{B}^{} = &TS_{B}+\sum_{i} {\mu_{i}^{}} {\rho_{i}^{}} - {p_{N}^{}} \nonumber\\
=&\sum_{i}2\int\frac{d^{3}\bm{k}}{(2\pi)^{3}} E^{*}_{i}(k)\Big{[} f(E^{*}_{i}(k))+\bar{f}(E^{*}_{i}(k)) \Big{]} \nonumber\\
&+\frac{1}{2}m_{\sigma}^{2}\sigma^{2}+\frac{1}{3}b\,m_{N}({g_{\sigma N}^{}} \sigma)^{3}
 +\frac{1}{4}c({g_{\sigma N}^{}} \sigma)^{4}\nonumber\\
&+\frac{1}{2}m^{2}_{\omega}
 \omega_{0}^{2}+\frac{1}{2}m^{2}_{\rho}
 \rho_{03}^{2}\,. \label{eq:energy}
\end{align}

In the mean field approximation, expected values of the meson fields are determined through their stationarity conditions, {\it i.e.},
\begin{align}
\frac{\partial \Omega_{B}}{\partial \sigma}=\frac{\partial
\Omega_{B}}{\partial \omega_{0}^{}}=\frac{\partial
\Omega_{B}}{\partial \rho_{03}^{}}=0,
\end{align}
which produce the following three equations of motion for mesons:
\begin{align}
m_{\sigma}^{2} \sigma = &
\sum_{i}g_{\sigma i}\rho^{s}_{i}-b\,m_{N}{g_{\sigma N}^{3}}\sigma^{2}
-c{g_{\sigma N}^{4}}\sigma^{3} \,, \label{eq:eommeson1}\\
{m_{\omega}^{2}} \omega_{0}^{} = &
\sum_{i}g_{\omega i}\rho_{i}^{}\,, \label{eq:eommeson2}\\
{m_{\rho}^{2}} {\rho_{03}^{}}  = &
\sum_{i}g_{\rho i}t_{3i}^{} \rho_{i}^{} \,, \label{eq:eommeson3}
\end{align}
with the scalar density for baryons given by
\begin{align}
\rho^{s}_{i}=2\int\frac{d^{3}\bm{k}}{(2\pi)^{3}}\frac{m^{*}_{i}}{E^{*}_{i}(k)} \Big{[} f(E^{*}_{i}(k))+\bar{f}(E^{*}_{i}(k)) \Big{]} \,.
\label{eq:scalardensity}
\end{align}

Now, it is left to specify the parameters in the RMF, which are fixed by fitting the properties of nuclear matter, including the saturation baryon number density $\rho_{0}^{}=0.16\:\mathrm{fm^{-3}}$, binding energy per nucleon $E/A=-16\:\mathrm{MeV}$, effective mass of nucleon at $\rho_{0}^{}$  $m_{N}^{*}=0.75\,m_{N}$ with $m_{N}=938\:\mathrm{MeV}$, the incompressibility $K=240\:\mathrm{MeV}$, and the symmetry energy $E_{s}=30.5\:\mathrm{MeV}$. The obtained parameters are given by $({g_{\sigma N}^{}}/m_{\sigma})^{2}\!=\!10.32\:\mathrm{fm^{2}}$, $({g_{\omega N}^{}}/m_{\omega})^{2}\!=\!5.41\:\mathrm{fm^{2}}$, $({g_{\rho N}^{}}/m_{\rho})^{2}\!=\!3.84\:\mathrm{fm^{2}}$, $b\!=\!6.97\times10^{-3}$ and $c\!=\!-4.85\times10^{-3}$, respectively. For the couplings between hyperons and mesons, we adopt the following relations: $g_{\sigma H}=0.7g_{\sigma N}$, $g_{\omega H}=0.783g_{\omega N}$ and $g_{\rho H}=0.783g_{\rho N}$, for more details, see {\it e.g.} Ref.~\cite{Glendenning:2000}.

To proceed,  we summarize some thermodynamical quantities relevant to electrons, neutrinos and photons, all of which are assumed to be non-interacting in our calculations. Furthermore, masses of electrons and neutrinos are also neglected. For the electrons, one has
\begin{align}
\rho_{e}^{} & =
\frac{1}{3\pi^{2}}\big(\mu_{e}^{3} + \pi^{2} {\mu_{e}^{}} T^{2}\big)\,,\\
\varepsilon_{e} & =
\frac{1}{4\pi^{2}}\big(\mu_{e}^{4}+2\pi^{2}\mu_{e}^{2}T^{2}+\frac{7\pi^{4}}{15}T^{4}\big)\,,\\
p_{e}^{} & =
\frac{1}{12\pi^{2}}\big(\mu_{e}^{4}+2\pi^{2}\mu_{e}^{2}T^{2}+\frac{7\pi^{4}}{15}T^{4}\big)\,,\\
S_{e} & =
\frac{T}{3}\big(\mu_{e}^{2}+\frac{7\pi^{2}}{15}T^{2}\big)\,,
\end{align}
which are the number density, energy density, pressure, and the entropy density, respectively, and $\mu_{e}^{}$ is the electron chemical potential.
For the electron neutrinos, one has
\begin{align}
\rho_{\nu_{e}}^{} & =
\frac{1}{6\pi^{2}}\big(\mu_{\nu_{e}}^{3}+\pi^{2}\mu_{\nu_{e}}T^{2}\big)\,, \\
\varepsilon_{\nu_{e}}^{} & =
\frac{1}{8\pi^{2}}\big(\mu_{\nu_{e}}^{4}+2\pi^{2}\mu_{\nu_{e}}^{2}T^{2}+\frac{7\pi^{4}}{15}T^{4}\big)\,, \\
p_{\nu_{e}}^{} & =
\frac{1}{24\pi^{2}}\big(\mu_{\nu_{e}}^{4}+2\pi^{2}\mu_{\nu_{e}}^{2}T^{2}+\frac{7\pi^{4}}{15}T^{4}\big)\,, \\
S_{\nu_{e}}^{} & =
\frac{T}{6}\big(\mu_{\nu_{e}}^{2}+\frac{7\pi^{2}}{15}T^{2}\big)  \, ,
\end{align}
with the chemical potential for electron neutrinos $\mu_{\nu_{e}}^{}$. Since the chemical potentials for $\mu$ and $\tau$ neutrinos are vanishing, we will not distinguish between them, and use $\nu_{x}^{}$ to denote them. The thermodynamical quantities relevant to $\nu_{x}^{}$ read
\begin{align}
\varepsilon_{\nu_{x}}= \frac{7\pi^{2}}{60}T^{4},\quad p_{\nu_{x}}=\frac{7\pi^{2}}{180}T^{4},\quad S_{\nu_{x}}  =  \frac{7\pi^{2}}{45}T^{3}\,.
\end{align}
At last, for photons we have
\begin{align}
\varepsilon_{\gamma}=\frac{\pi^{2}}{15}T^{4}, \quad
p_{\gamma}^{}=\frac{\pi^{2}}{45}T^{4},   \quad
S_{\gamma}=\frac{4\pi^{2}}{45}T^{3} \, .
\end{align}

Summing up all the contributions of the components, we obtain the energy density, pressure, and the entropy density of the stellar matter of a newly born NS, as given by
\begin{align}
\varepsilon & =\varepsilon_{B}^{} +\varepsilon_{e}^{} +\varepsilon_{\nu_{e}}^{} +\varepsilon_{\nu_{x}}^{} +\varepsilon_{\gamma}^{}  \, ,\\
p & = p_{B}^{} + p_{e}^{} + p_{\nu_{e}}^{} +p_{\nu_{x}}^{} +p_{\gamma}^{} \,,\\
S' & =S_{B}^{} + S_{e}^{} +S_{\nu_{e}}^{} +S_{\nu_{x}}^{} +S_{\gamma}^{} \,,\label{eq:entropy}
\end{align}
respectively, where we save notation $S$ for other use in the following.
Because of the high temperature and density inside a newly born NS, the timescale of arriving at a beta equilibrium is of order of $\sim 10^{-8}\,\mathrm{s}$~\cite{Fu:2008zzg}, being much smaller than the period of the nonradial oscillations about $10^{-4}\sim 10^{-2}\,\mathrm{s}$. Thus, it is reasonable to assume that the stellar matter is always in beta equilibrium, which entails that the chemical potentials for the baryon octets read
\begin{align}
\mu_{i}=\mu_{n}-Q_i(\mu_{e}-\mu_{\nu_e})\, ,\label{eq:beta}
\end{align}
with electric charge $Q_{i}^{}$ for baryon $i$. Therefore, we have three independent chemical potentials, {\it i.e.}, $\mu_{n}^{}$, $\mu_{e}^{}$, and $\mu_{\nu_{e}}^{}$. Given a baryon number density $\rho_{B}^{}=\sum_{i} \rho_{i}^{} $, an electron abundance $Y_{e}^{}$ or a lepton abundance $Y_{L}^{}$,
and an entropy per baryon $S=S'/{\rho_{B}^{}}$, combined with the constraint that the stellar matter is electric charge free, {\it i.e.}, $\sum_{j}Q_{j}^{} {\rho_{j}^{}}=0$ with $j$ running over baryons and leptons, one can obtain the three independent chemical potentials and all other thermodynamical quantities mentioned above, by solving Eqs.~(\ref{eq:eommeson1})--(\ref{eq:eommeson3}) and (\ref{eq:entropy}).

\section{\label{sec:MIT}Newly Born Strange Quark Stars---in MIT Bag Model}

In this section we employ the simplest model of the strange quark matter, {\it i.e.}, the MIT bag model~\cite{Chodos:1974je,Farhi:1984qu}, to describe the EOS of the matter in a newly born SQS.
A more sophisticated Nambu--Jona-Lasinio model description, which incorporate the information of the dynamical chiral symmetry breaking of QCD, will be delayed to Sec.\ref{sec:NJL}.
Newly born SQS matter is usually assumed to be composed of three flavor quarks ($u$, $d$, and $s$), electrons, neutrinos, thermal photons and gluons. Parameters in the bag model are chosen to be $m_{u}=m_{d}=0$, $m_{s}=150\,\mathrm{MeV}$, and a bag constant $B^{1/4}=154.5\,\mathrm{MeV}$, which, for the SQS matter, produce a mass per baryon of $928\,\mathrm{MeV}$, being slightly smaller than that $931\,\mathrm{MeV}$ in stablest nucleus ${^{56}}\mathrm{Fe}$, that is consistent with the conjecture on the strange quark matter~\cite{Witten:1984rs}. For more discussions about the MIT bag model, see, {\it e.g.}, Ref.~\cite{Farhi:1984qu}.

To proceed, we present some thermodynamical quantities relevant to the matter of newly born SQSs. For the $u$ and $d$ quarks, since they are relativistic, {\it i.e.}, massless, one has
\begin{align}
\rho_{u}^{} & =
\frac{1}{\pi^{2}}\big(\mu_{u}^{3}+\pi^{2} {\mu_{u}^{}} T^{2}\big)\,,\\
\varepsilon_{u} & = \frac{3}{4\pi^{2}}\big(\mu_{u}^{4}+2\pi^{2}\mu_{u}^{2}T^{2}+\frac{7\pi^{4}}{15}T^{4}\big)\,,\\
p_{u}^{} & = \frac{1}{4\pi^{2}}\big(\mu_{u}^{4}+2\pi^{2}\mu_{u}^{2}T^{2}+\frac{7\pi^{4}}{15}T^{4}\big)\,,\\
S_{u} & =  T\big(\mu_{u}^{2}+\frac{7\pi^{2}}{15}T^{2}\big)\, .
\end{align}
Same expressions also apply to $d$ quarks.
Notations in the equations above are the same as those in Sec.\ref{sec:NSs}. For the massive strange quark, we have
\begin{align}
\rho_{s}^{} & =  6\int\frac{d^{3}\bm{k}}{(2\pi)^{3}}
\big{[} f(E_{s}(k))-\bar{f}(E_{s}(k)) \big{]} \,,\\
\varepsilon_{s}^{} & =  6\int\frac{d^{3}\bm{k}}{(2\pi)^{3}}
E_{s}(k) \big{[} f(E_{s}(k))+\bar{f}(E_{s}(k)) \big{]} \,,\\
p_{s}^{} & =  6\int\frac{d^{3}\bm{k}}{(2\pi)^{3}}
\frac{k}{3}\frac{d E_{s}(k)}{d k} \big{[} f(E_{s}(k))+\bar{f}(E_{s}(k)) \big{]} \,,\\
S_{s} & =  \frac{6}{T}\int\frac{d^{3}\bm{k}}{(2\pi)^{3}}
\Big{[} \Big{(} E_{s}(k)+\frac{k}{3}\frac{d E_{s}(k)}{d
k}-\mu_{s} \Big{)} f(E_{s}(k))\nonumber\\
& \quad +\Big{(} E_{s}(k)+\frac{k}{3}\frac{d E_{s}(k)}{d
k} + {\mu_{s}^{}} \Big{)} \bar{f}(E_{s}(k)) \Big{]}   \, ,
\end{align}
with $E_{s}(k)=(k^{2}+m_{s}^{2})^{1/2}$ and the fermionic distribution function given in Eqs.~(\ref{eq:fermi}) and (\ref{eq:antifermi}), but with the chemical potential replaced by that of strange quarks. Thermodynamics relevant to electrons, neutrinos, and thermal photons have already presented in the last section, and here what we need in addition are those related to gluons, which read
\begin{align}
\varepsilon_{g}=\frac{8\pi^{2}}{15}T^{4},   \quad
p_{g}^{} =\frac{8\pi^{2}}{45}T^{4},     \quad
S_{g}=\frac{32\pi^{2}}{45}T^{3},
\end{align}
where interactions among gluons are neglected as well. Finally, one obtains the thermodynamical quantities describing the stellar matter of a newly born SQS as
\begin{align}
\varepsilon &=\Big(\sum_{i=u,d,s}\varepsilon_{i}\Big)+\varepsilon_{e}+\varepsilon_{\nu_{e}}
+\varepsilon_{\nu_{x}}+\varepsilon_{\gamma}+\varepsilon_{g}+B  \, ,  \\
p & =\Big(\sum_{i=u,d,s}{p_{i}^{}}\Big) + p_{e}^{} +p_{\nu_{e}}^{} +p_{\nu_{x}}^{} +p_{\gamma}^{} +p_{g}^{} - B  \,  ,\\
S^{\prime} & =\Big(\sum_{i=u,d,s}S_{i}\Big)+S_{e}+S_{\nu_{e}}+S_{\nu_{x}}+S_{\gamma}+S_{g}\,,\label{eq:SMIT}
\end{align}
where $B$ is the bag constant. Similar with the stellar matter of a newly born NS, strange quark matter in a newly born SQS is also in beta equilibrium, leading to the relations for the chemical potentials:
\begin{align}
\mu_{e}^{} +\mu_{u}^{} & =  \mu_{d}^{} + \mu_{\nu_{e}}^{}  \,  ,  \label{eq:betaMIT1}\\
\mu_{s}^{} & = \mu_{d}^{}  \,.\label{eq:betaMIT2}
\end{align}
Like in the NS case, in order to determine the equilibrium configuration of a SQS, one also needs several conservation conditions, such as the baryon number $(\rho_{u}^{} +\rho_{d}^{} +\rho_{s}^{})/3=\rho_{B}^{}$, lepton number density abundance $Y_{e}^{}+Y_{\nu_{e}}^{}=Y_{L}^{}$, and the electric charge neutral condition
\begin{align}
\frac{2}{3}\rho_{u}^{} - \frac{1}{3}\rho_{d}^{} - \frac{1}{3}\rho_{s}^{} -\rho_{e}^{} =0
\, .  \label{eq:neutralMIT}
\end{align}

\section{\label{sec:numerical}Numerical Results and Discussions}

As we have discussed above, neutrinos are trapped in a newly born NS or SQS, and thus the adiabatic sound speed in Eq.~(\ref{eq:cs}) can be rewritten as
\begin{align}
c_{s}^{2}=\left(\frac{\partial p}{\partial \rho}\right)_{S,\,Y_{\mathrm{L}}}   \,  .   \label{eq:cs2}
\end{align}
It might has been noticed that, to determine $c_{s}$ as well as the equilibrium sound speed $c_{e}$ in Eq.~(\ref{eq:ce}), equilibrium configuration of a star, {\it i.e.}, the dependence of $\rho_{B}^{}$, $S$, and $Y_{L}$ or $Y_{e}$ on the radial coordinate, has to be provided. Following Ref.~\cite{Fu:2008bu}, we employ the equilibrium configurations of newly born NSs, obtained in 2D hydrodynamic simulations of core-collapse supernovae by the Arizona Group~\cite{Dessart:2005ck}. In our calculations we choose three representative instants of time, {\it i.e.}, $t=100$, $200$ and $300\,\mathrm{ms}$, after the core bounces in the core-collapse supernovae. Since the eigenfrequencies of $g$-mode vary only a little during the first second after the core bounces, for instance, it has been found that the eigenfrequencies resides in a narrow range of $727\,\mathrm{Hz}\sim 819\,\mathrm{Hz}$, from $0.3\,\mathrm{s}$ to $1\,\mathrm{s}$ in a newly born NS model~\cite{Ferrari:2002ut}, we restrict computations in this work within $300\,\mathrm{ms}$ for simplicity.

We integrate Eqs.~(\ref{eq:Osc1}) and (\ref{eq:Osc2}) from the center to a radius of about $20\,\mathrm{km}$, where convective instabilities set in~\cite{Dessart:2005ck}. The stellar mass of the newly born NS inside the radius of $20\,\mathrm{km}$ is chosen to be $0.8\,\mathrm{M}_{\odot}$, $0.95\,\mathrm{M}_{\odot}$ and $1.05\,\mathrm{M}_{\odot}$ at $t\!=\!100$, $200$ and $300\,\mathrm{ms}$, respectively, in agreement with the supernovae simulations~\cite{Dessart:2005ck}. The mass increases with time, because mantle materials of the progenitor star are accreted onto the newly born NS continuously.

For newly born SQSs, following Ref.~\cite{Fu:2008bu} we assume that the dependence of the entropy $S$ and lepton abundance $Y_{L}$ on the radial coordinate for newly born SQSs, is the same as for NSs. In comparison with NSs, a newly born SQS has a smaller size, with a radius of about $10\,\mathrm{km}$ where the pressure vanishes. The stellar mass of the SQS in our calculations is chosen to be the same as the NS, for the three different instants after the core bounces.

\begin{table}[!t]
\begin{center}
\caption{Eigenfrequencies (Hz) of quadrupole ($l=2$) oscillations of $g$-mode for newly born NSs and SQSs at three instants of time (ms) after the core bounces (Quoted from Table~I in Ref.~\cite{Fu:2008bu}).}
\label{tab:gmode}
\begin{tabular}{L|ccc|ccc}
\hline \hline                    \vspace{0.1cm}
Radial order of $g$-mode & \multicolumn{3}{c}{Neutron Star} & \multicolumn{3}{|c}{Strange Quark Star}  \\\hline
 & $t\!=\!100$&$t\!=\!200$&$t\!=\!300$&$t\!=\!100$&$t\!=\!200$&$t\!=\!300$\\
 $n=1$ & 717.6& 774.6 & 780.3 &  82.3 & 78.0 & 63.1 \\
 $n=2$ & 443.5& 467.3 & 464.2 &  52.6 & 45.5 & 40.0   \\
 $n=3$ & 323.8& 339.0 & 337.5 &  35.3 & 30.8 & 27.8   \\
\hline \hline
\end{tabular}
\end{center}
\end{table}

\begin{figure}[t]
\includegraphics[scale=0.58]{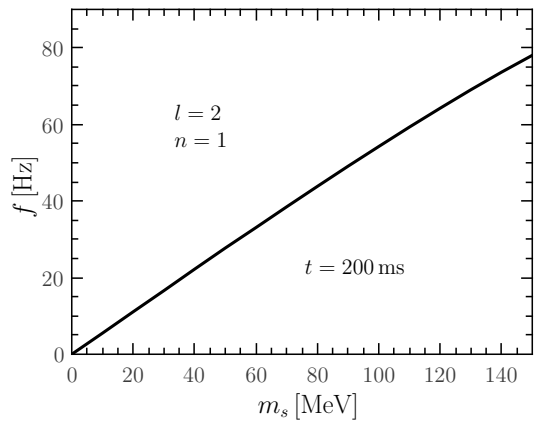}
\caption{Eigenfrequency of the $g$-mode for a SQS with $l=2$ and $n=1$, as a function of the strange quark mass $m_s$, where we choose $t\!=\!
200\mathrm{ms}$ illustratively.}\label{fig:fms}
\end{figure}

In this work, for newly born NSs, we extend our former calculations in Ref.~\cite{Fu:2008bu} to include the hyperon degrees of freedom. However, since the stellar masses are all just about one solar mass for the three instants of time, which results in that the baryon densities in the stellar center are less than $2\rho_{0}^{}$, we find then that the hyperons do not appear in the newly born NSs for all the three instants. Thus, the eigenfrequencies of the $g$-mode oscillations for newly born NSs calculated in this work are identical to those presented in the Table I in Ref.~\cite{Fu:2008bu}. We quote the result here in \tab{tab:gmode} for the convenience of discussion. Note that $g$-mode quadrupole pulsations of compact stars with low radial orders, especially $n=1$, are significantly potential sources of gravitational waves~\cite{Misner:1970,Owen:2005fn,Lai:2006pr}, which are hopefully to be detected in the near future by such as Advanced LIGO. One can see from \tab{tab:gmode} that, for the $g$-mode quadrupole oscillations of NSs with $n=1$, the eigenfrequencies are $717.6$, $774.6$ and $780.3\,\mathrm{Hz}$, respectively, for the three different instants. These results are consistent with supernovae simulations~\cite{Burrows:2006ci,Ott:2006qp} as well as those computed from general relativity~\cite{Ferrari:2002ut}. In contrast to the case of NSs, eigenfrequencies of the $g$-mode for SQSs with $n=1$, $l=2$, at the three representative instants after the core bounces, are $82.3$, $78.0$ and $63.1\,\mathrm{Hz}$ respectively, which are lower by an order of magnitude. Table~\ref{tab:gmode} also shows the results for higher radial orders with $n>1$. One can still find that eigenfrequencies of NSs are larger than those of SQSs by an order of magnitude. It was found in Ref.~\cite{Fu:2008bu} that this difference arises from the fact that nucleons, the major components of the stellar matter of NSs, are massive and non-relativistic, whereas SQSs are composed of relativistic particles. In fact, nonvanishing $g$-mode eigenfrequencies of a SQS are attributed to a finite strange quark mass $m_s$, since the masses of $u$ and $d$ quarks are negligible. Therefore, if we deal with $m_{s}^{}$ as a parameter, and reduce it from a finite value to zero artificially, one would have expected that the $g$-mode eigenfrequency of a SQS decreases with $m_{s}^{}$, and finally vanishes when $m_{s}^{}=0$. This expectation is confirmed by the numerical result shown in \Fig{fig:fms}, where we have chosen the instant $t\!=\!200\mathrm{ms}$ for an illustrative purpose. Note that the mass of strange quark is much smaller than that of nucleons, which is the reason why the eigenfrequencies of $g$-mode oscillations of a SQS are much smaller than those of a NS.

%
%%%%%%%%%%%%%%%%%%%%%%%%%%%%%
%%%%%%%%%%%%%%%%%%%%%%%%%
%

%
%%%%%%%%%%%%%%%%%%%%%%%%%%%%%
\begin{figure*}[t]
\includegraphics[scale=0.45]{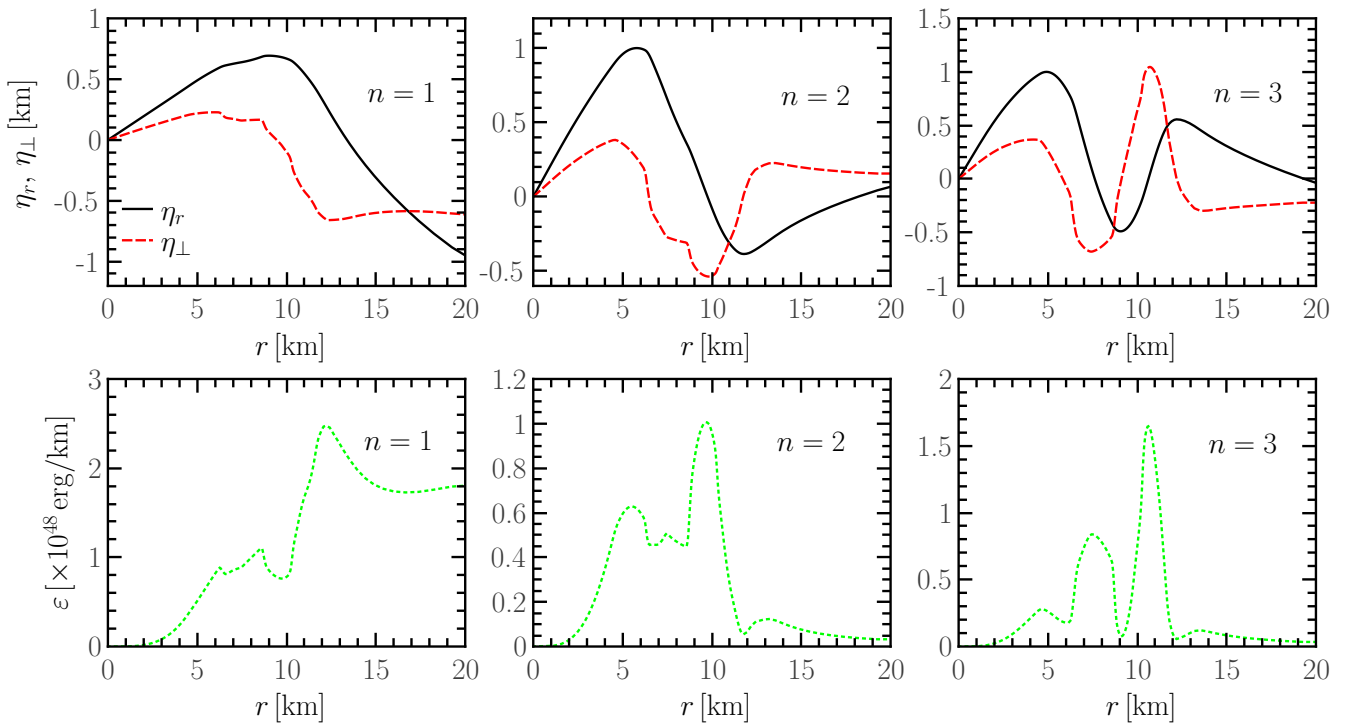}
\caption{Upper panels: radial $\eta_{r}^{}$ and nonradial $\eta_{\bot}^{}$ components of the displacement in \Eq{eq:eta} as functions of the radial coordinate $r$, for the $g$-mode eigen-oscillations of a newly born NS with $l=2$, and $n=1,\,2,\,3$, respectively. $t\!=\!200\,\mathrm{ms}$ is chosen as a representative instant of time. Lower panels: corresponding energy of stellar oscillations per unit of the radial coordinate as a function of $r$.}\label{fig:waveformgNS}
\end{figure*}
%%%%%%%%%%%%%%%%%%%%%%%%%
%
%
%%%%%%%%%%%%%%%%%%%%%%%%%%%%%
\begin{figure*}[htb]
\begin{center}
\includegraphics[scale=0.45]{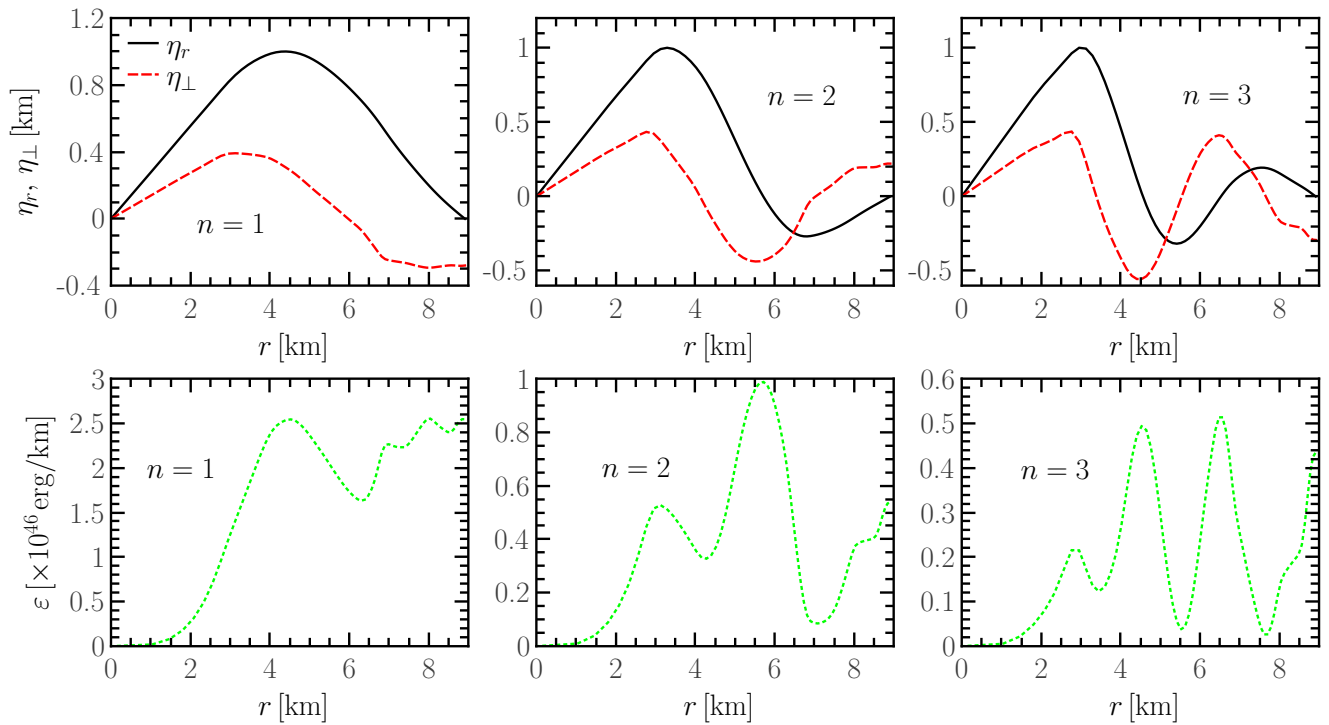}
\caption{Same as \Fig{fig:waveformgNS}, but for a newly born SQS.}\label{fig:waveformgSQS}
\end{center}
\end{figure*}
%%%%%%%%%%%%%%%%%%%%%%%%%%

In \Fig{fig:waveformgNS} and \Fig{fig:waveformgSQS} we show the eigenfunctions of the quadrupole oscillations of $g$-mode, described by the components of the displacement $\eta_{r}^{}$ and $\eta_{\bot}^{}$ as functions of the radial coordinate, for the newly born NS and SQS, respectively. Here we choose the radial order $n=1,\,2,\,3$, and $t\!=\!200\,\mathrm{ms}$ after the core bounce as a representative instant. Note that the amplitude of oscillations is normalized so that the maximal value of $\eta_{r}^{}$ is $1\,\mathrm{km}$. Furthermore, the energy of the stellar oscillation of an eigen-mode is given by \cite{Reisenegger:1992}
\begin{equation}
E=\frac{\omega^{2}}{2}\int_{0}^{R}\rho
r^{2} \big{[} \eta_{r}^{2}+l(l+1)\eta_{\bot}^{2} \big{]} d r\, .\label{mode_E}
\end{equation}
In \Fig{fig:waveformgNS} and \Fig{fig:waveformgSQS} we also show the oscillation energy per unit of the radial coordinate, viz.
\begin{equation}
\varepsilon=\frac{\omega^{2}}{2}\rho
r^{2} \big{[} \eta_{r}^{2}+l(l+1)\eta_{\bot}^{2} \big{]} \, ,\label{E_density}
\end{equation}
for the $g$-mode oscillations of the newly born NS and SQS, respectively. We find that the energy density $\varepsilon$ in \Eq{E_density} for the SQS is lower than that for the NS by two orders of magnitude, because of the big difference of eigenfrequencies.

%
%%%%%%%%%%%%%%%%%%%%%%%%%%%%%
\begin{figure*}[t]
\includegraphics[scale=0.45]{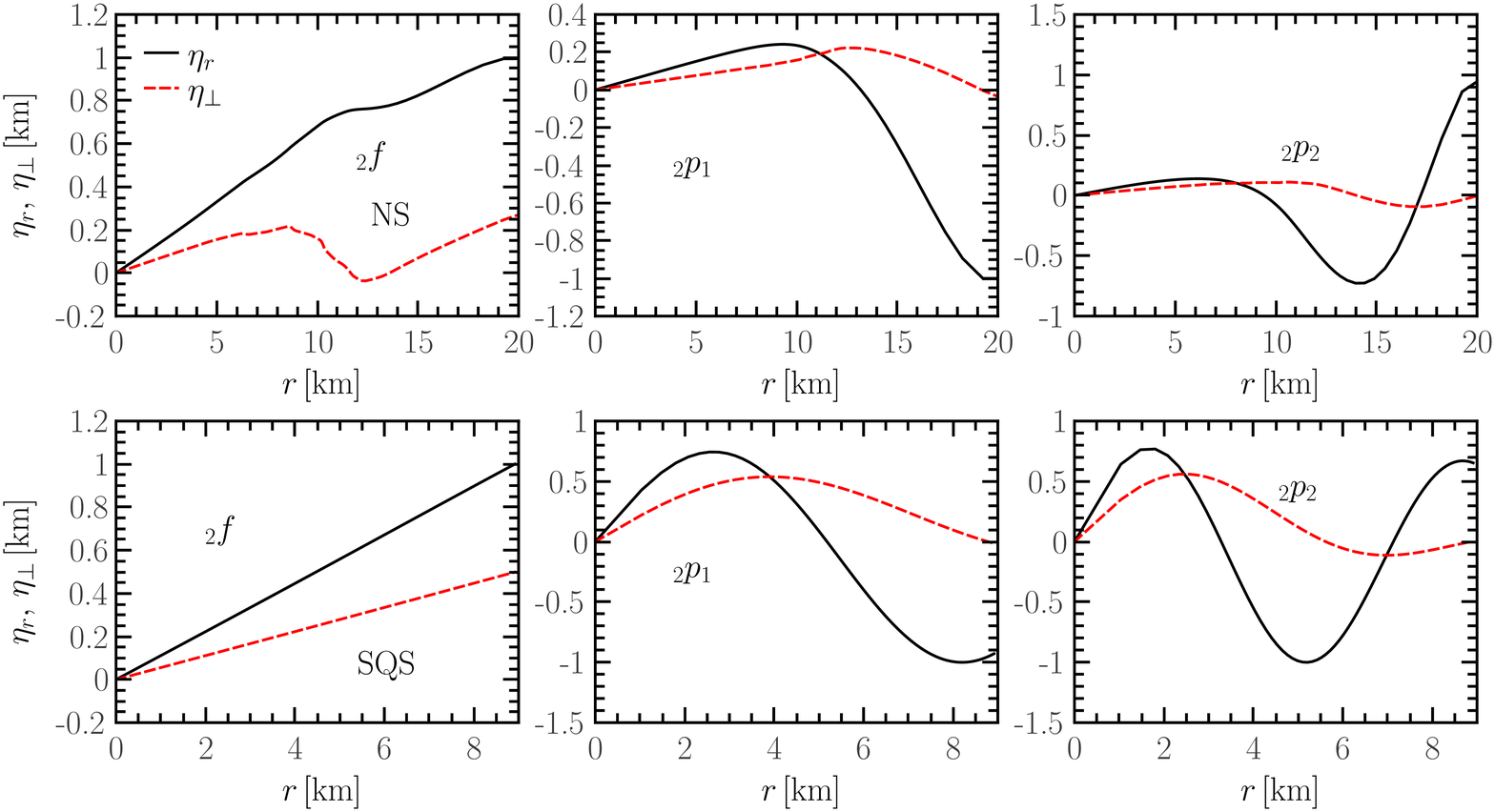}
\caption{$\eta_{r}^{}$ and $\eta_{\bot}^{}$ as functions of the radial coordinate, for the $f$- and $p$-modes of a newly born NS (upper panels) and SQS (lower panels) with $l=2$. $t\!=\!200\,\mathrm{ms}$ is chosen as a representative instant of time.}\label{fig:waveformfp}
\end{figure*}
%%%%%%%%%%%%%%%%%%%%%%%%%
%

Up to now, we have only discussed the stellar nonradial oscillations of $g$-mode. In fact, besides the gravitational mode, there are other modes for nonradial oscillations, such as the $f$-mode, {\it i.e.}, the mode with the radial order $n=0$, and the $p$-mode, both of which can also be obtained by solving the nonradial oscillation equations in \Eq{eq:Osc1} and \Eq{eq:Osc2}. In this work we would like to present some results for the $f$- and $p$-modes, in comparison with the relevant ones of $g$-mode. In the following we employ $_{l}p_{n}$ to denote the pressure mode with angular index $l$ and the radial order $n$, and $_{l}f$ the fundamental mode with angular index $l$.

\begin{table}[htb]
\begin{center}
\caption{Eigenfrequencies (Hz) of quadrupole ($l=2$) oscillations of $f$- and $p$-modes for newly born NSs and SQSs at three instants of time (ms) after the core bounces.}
\label{tab:fpmode}
\begin{tabular}{S|ccc|ccc}
\hline \hline                    \vspace{0.1cm}
Modes &\multicolumn{3}{c}{Neutron Star} & \multicolumn{3}{|c}{Strange Quark Star}  \\
\hline
 & $t\!=\!100$&$t\!=\!200$&$t\!=\!300$&$t\!=\!100$&$t\!=\!200$&$t\!=\!300$\\
 $_{2}f$     & 1103& 1133 & 1176 &  2980  & 2997  & 3016 \\
 $_{2}p_{1}$ & 2265& 2426 & 2494 &  18282 & 17330 & 16792   \\
 $_{2}p_{2}$ & 3780& 4054 & 4179 &  28792 & 27288 & 26438   \\
 $_{2}p_{3}$ & 5319& 5702 & 5869 &  38988 & 36950 & 35798   \\
\hline \hline
\end{tabular}
\end{center}
\end{table}

In \tab{tab:fpmode} we show our obtained eigenfrequencies of quadrupole oscillations of $f$- and $p$-modes for newly born NSs and SQSs at the three representative instants. One can observe evidently that the eigenfrequencies of $f$- and $p$-modes are larger than those of the $g$-mode for both NSs and SQSs, which is because they have different origins. Oscillations of $g$-mode arise from the buoyancy inside the star, but those of $f$- and $p$-modes result from the pressure.
As discussed in detail in Ref.~\cite{Fu:2008bu}, the eigenfrequencies of $g$-mode are closely related with the difference of the equilibrium and adiabatic sound speeds, {\it i.e.}, the Brunt-V\"{a}is\"{a}l\"{a} frequency in \Eq{eq:BV}, while those of $f$- and $p$-modes are connected with the value of sound speeds. The difference between the two sound speeds are much smaller than each of them, thus the eigenfrequencies of $g$-mode are smaller than those of $f$- and $p$-modes. In \tab{tab:fpmode} we also find that, for the $f$- and $p$-modes, eigenfrequencies of SQSs are larger than those of NSs, in contradistinction to the $g$-mode. This is because the SQSs are more compact than NSs, and specifically densities in the outer region of the NSs are significantly low, which leads to smaller sound speeds there. In \Fig{fig:waveformfp} we show the eigenfunctions of several $f$- and $p$-modes for the NS and SQS.

\section{\label{sec:NJL}Newly Born Strange Quark Stars---in NJL Model}

In \sec{sec:MIT} we employ the MIT bag model to describe the quark matter and construct the configuration of the SQS. Nonperturbative QCD is characteristic of the dynamical chiral symmetry breaking (DCSB) at low energy. Since there is no explicit demonstration of the DCSB in the bag model, it can not describe the important feature of QCD. In this section we would like to adopt the Nambu--Jona-Lasinio (NJL) model \cite{Nambu:1961tp,Nambu:1961fr}, which possesses the chiral symmetry and DCSB, for more details about the NJL model and applications of the model in hadron physics and QCD phase diagram, see, {\it e.g.}, Refs.~\cite{Klevansky:1992qe,Hatsuda:1994pi,Buballa:2003qv}. The Lagrangian density for the 2+1 flavour NJL model is given by~\cite{Rehberg:1995kh}
\begin{align}
\mathcal{L}=&\, \bar{\psi}\left(i\gamma_{\mu}\partial^{\mu} \!- \! \hat{m}_{0}\right)\psi +G\sum_{a=0}^{8}\big[\left(\bar{\psi}\tau_{a}\psi\right)^{2}
 \! + \! \left(\bar{\psi}i\gamma_{5}\tau_{a}\psi\right)^{2}\big]\notag   \nonumber \\
&-K\left[\textrm{det}_{f}\left(\bar{\psi}\left(1+\gamma_{5}\right)\psi\right)
 +\textrm{det}_{f}\left(\bar{\psi}\left(1-\gamma_{5}\right)\psi\right)\right]
 ,\label{eq:NJLlagragian}
\end{align}
with the quark fields $\psi=(\psi_{u},\psi_{d},\psi_{s})^{T}$, the matrix of the current quark masses $\hat{m}_{0}=\textrm{diag}(m_{u},m_{d},m_{s})$. We choose $m_{l}^{}:=m_{u}^{}=m_{d}^{}$ and $m_{l}^{}<m_{s}^{}$ as same as the case in \sec{sec:MIT}. The four fermion interactions in \Eq{eq:NJLlagragian} is symmetric under the transformations of $U(3)_{L}\otimes U(3)_{R}$ in the flavour space, with the interaction strength $G$. $\tau_{0}^{}=\sqrt{\frac{2}{3}} \mathbf{1}_{f}$ and Gell-Mann matrices $\tau_{i}^{}\,(i=1,\ldots,8)$ are normalized such that we have $\mathrm{tr}(\tau_{a}\tau_{b})=2\delta_{ab}$. The 't Hooft interactions in \Eq{eq:NJLlagragian}, with the coupling strength $K$, break the symmetry of $U_{A}(1)$, but keep that of $SU(3)_{L}\otimes SU(3)_{R}$. The parameters in the model read: the current mass of light quarks $m_{l}=5.5\;\mathrm{MeV}$, the strange quark mass $m_{s}=140.7\;\mathrm{MeV}$, coupling strengths $G\Lambda^{2}=1.835$ and $K\Lambda^{5}=12.36$, with the UV cutoff $\Lambda=602.3\;\mathrm{MeV}$. These parameters are fixed by fitting observables in vacuum, including $\pi$ meson mass $m_{\pi}=135.0\;\mathrm{MeV}$, $K$ meson mass $m_{K}=497.7\;\mathrm{MeV}$, $\eta^{\prime}$ meson mass $m_{\eta^{\prime}}=957.8\;\mathrm{MeV}$ , and the $\pi$ decay constant $f_{\pi}=92.4\;\mathrm{MeV}$, for more details, see, {\it e.g.}, Ref.~\cite{Rehberg:1995kh}.

The thermodynamical potential density in the mean field approximation reads
\begin{align}
\Omega_{\mathrm{QM}}=&-2N_{c}\!\!\sum_{i=u,d,s}\int\frac{d^{3}\bm{k}}{(2\pi)^{3}}\bigg\{
\beta^{-1}\ln\Big[1+e^{-\beta(E_{i}^{}(k) - \mu_{i}^{})}\Big]\nonumber\\
&+\beta^{-1}\ln\Big[1+e^{-\beta(E_{i}^{}(k) + \mu_{i}^{})}\Big]\,
+E_{i}^{} \bigg\}\nonumber\\
&+2G\left({\phi_{u}}^{2} +{\phi_{d}}^{2}+{\phi_{s}}^{2}\right) - 4K \phi_{u}^{}\,\phi_{d}^{}\,\phi_{s}^{}+C\,,
\label{eq:NJLpoten}
\end{align}
where $C$ is a constant, to be determined in the following.
The subscript $_{\mathrm{QM}}$ denotes quark matter and $N_{c}^{}=3$ is the number of colours.
The dispersion relation for quarks is given by
\begin{equation}
E_{i}^{}(k)=\sqrt{k^{2}+M_{i}^{2}}\,,
\end{equation}
with the constituent quark mass of flavour $i$, {\it i.e.}, $M_{i}^{}$ being
\begin{equation}
M_{i}^{}=m_{0}^{i} - 4G\phi_{i}^{} + 2K\phi_{j}^{}\,\phi_{k}^{},\label{eq:constituentmass}
\end{equation}
where $\phi_{i}^{}$ is the quark condensate of flavour $i$, reading
\begin{align}
\phi_{i}^{} = &-2N_{c}\int\frac{d^{3}\bm{k}}{(2\pi)^{3}}\frac{M_{i}}{E_{i}}
\Big{[} 1-f\big(E_{i}(k)\big) - \bar{f}\big(E_{i}(k)\big) \Big{]}, \label{eq:condensate}
\end{align}
where $f$ and $\bar{f}$ are the fermionic distribution functions in Eq.~\eq{eq:fermi} and \eq{eq:antifermi} with $E^{*}$ and $\mu^{*}$ replaced with $E$ and $\mu$, respectively. In the same way, employing the thermodynamical potential in \Eq{eq:NJLpoten}, one can obtain other thermodynamical quantities, such as the quark number density reading
\begin{align}
\rho_{i}^{} =&-\frac{\partial \Omega_{\mathrm{QM}}}{\partial \mu_{i}}\nonumber\\
=&\, 2N_{c}\int\frac{d^{3}\bm{k}}{(2\pi)^{3}} \Big{[} f\big(E_{i}(k)\big) - \bar{f}\big(E_{i}(k)\big) \Big{]}\,,
\label{eq:NJLdensity}
\end{align}
the entropy density
\begin{align}
S_{\mathrm{QM}}=&-\frac{\partial \Omega_{\mathrm{QM}}}{\partial T}\nonumber\\
=&\, \frac{2N_{c}}{T}\sum_{i=u,d,s}\int\frac{d^{3}\bm{k}}{(2\pi)^{3}} \Big{[} \Big(E_{i}(k) +\frac{k^{2}}{3E_{i}(k)}-\mu_{i}\Big)\nonumber\\
&\times f\big(E_{i}(k)\big) + \Big(E_{i}(k)+\frac{k^{2}}{3E_{i}(k)}+\mu_{i}\Big)\bar{f}\big(E_{i}(k)\big)\Big{]} \,,
\label{eq:NJLentropy}
\end{align}
the pressure
\begin{align}
p_{\mathrm{QM}}^{} = &-\Omega_{\mathrm{QM}}^{}     \nonumber\\
=&\, 2N_{c}\sum_{i=u,d,s}\int\frac{d^{3}\bm{k}}{(2\pi)^{3}}
\frac{k^{2}}{3E_{i}(k)} \Big{[} f\big(E_{i}(k)\big) \! + \! \bar{f}\big(E_{i}(k)\big) \Big{]} \nonumber\\
&+2N_{c}\sum_{i=u,d,s}\int\frac{d^{3}\bm{k}}{(2\pi)^{3}}E_{i}(k)\nonumber\\
&-2G\left({\phi_{u}}^{2}
+{\phi_{d}}^{2}+{\phi_{s}}^{2}\right)+4K\phi_{u}\,\phi_{d}\,\phi_{s}-C\,, \label{eq:NJLpressure}
\end{align}
from which we can define a effective bag constant as
\begin{align}
B_{\textrm{eff}}=&-2N_{c}\sum_{i=u,d,s}\int\frac{d^{3}\bm{k}}{(2\pi)^{3}}E_{i}(k)\nonumber\\
&+2G\left({\phi_{u}}^{2}
+{\phi_{d}}^{2}+{\phi_{s}}^{2}\right)-4K\phi_{u}\,\phi_{d}\,\phi_{s}+C\,. \label{eq:Beff}
\end{align}
In contrast to the bag constant in the MIT bag model, $B_{\textrm{eff}}$ in \Eq{eq:Beff} is dependent on the quark condensates, and thus on the density, rather than an absolute constant. Finally, one obtains the energy density as given by
\begin{align}
\varepsilon_{\mathrm{QM}}^{}=&\, TS_{\mathrm{QM}}+\sum_{i=u,d,s}\mu_{i}\rho_{i} -p_{\mathrm{QM}}^{} \nonumber\\
=& \, 2N_{c}\sum_{i=u,d,s}\int \!\! \frac{d^{3}\bm{k}}{(2\pi)^{3}} E_{i}^{}(k)
\Big{[} f\big(E_{i}^{}(k)\big)\! +\bar{f}\big(E_{i}^{}(k)\big) \Big{]} \nonumber\\
&+B_{\textrm{eff}}\,. \label{eq:NJLenergy}
\end{align}
Note that the constant $C$ in \Eq{eq:NJLpoten} is determined with $\varepsilon_{\mathrm{QM}}^{}= -p_{\mathrm{QM}}^{} =0$ in the vacuum.

%
%%%%%%%%%%%%%%%%%%%%%%%%%%%%%
\begin{figure}[t]
\begin{center}
\includegraphics[scale=0.6]{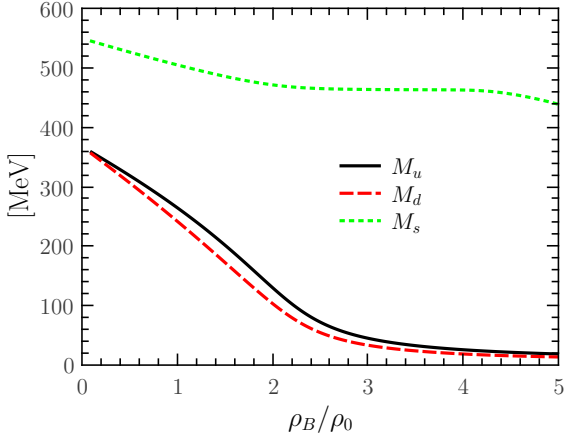}
\caption{Constituent quark masses as functions of the baryon number density, calculated in the NJL model, for the charge-neutral quark matter at zero temperature and neutrino abundance.}\label{fig:quarkmass}
\end{center}
\end{figure}
%%%%%%%%%%%%%%%%%%%%%%%%%
%

%
%%%%%%%%%%%%%%%%%%%%%%%%%%%%%
\begin{figure*}[htb]
\begin{center}
\includegraphics[scale=0.6]{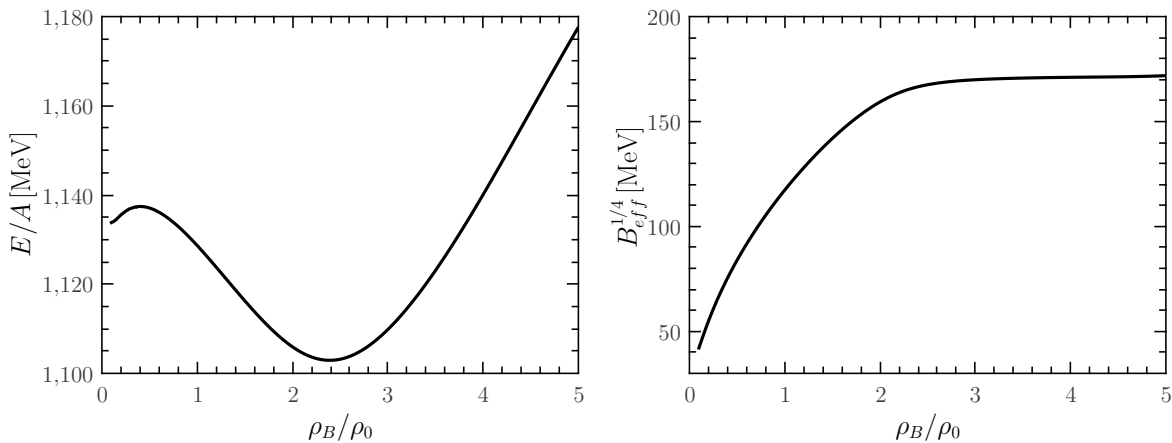}
\caption{Energy per baryon number $E/A$ (left panel) and the effective bag constant $B_{\textrm{eff}}^{1/4}$ defined in \Eq{eq:Beff} (right panel) as functions of the baryon number density for the charge-neutral quark matter at zero temperature and neutrino abundance.}\label{fig:EAB}
\end{center}
\end{figure*}
%%%%%%%%%%%%%%%%%%%%%%%%%
%

%
%%%%%%%%%%%%%%%%%%%%%%%%%%%%%
\begin{figure*}[htb]
\includegraphics[scale=0.6]{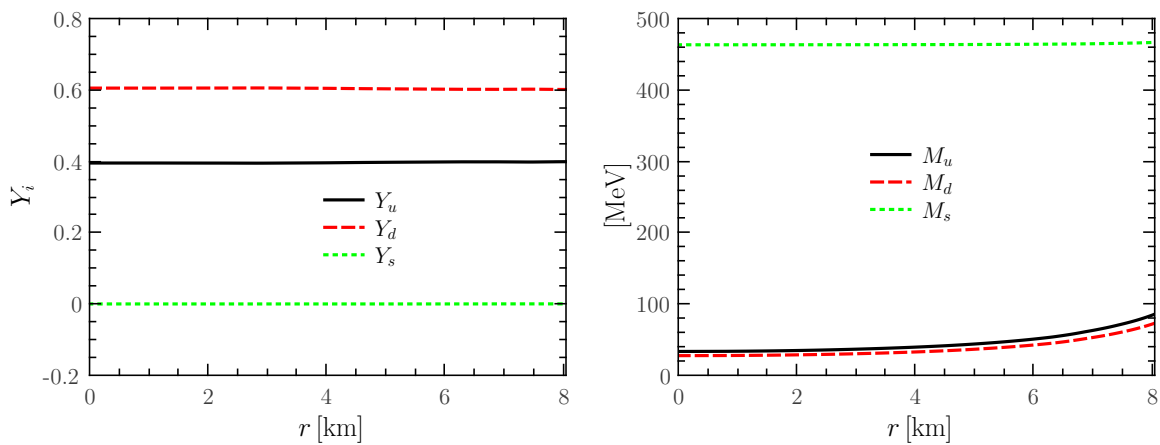}
\caption{Abundances of the three flavour quarks (left panel) and their corresponding constituent quark masses (right panel) as functions of the stellar radial coordinate for a newly born SQS. $t\!=\!200\,\mathrm{ms}$ is chosen as a representative instant of time.}\label{fig:Yimass}
\end{figure*}
%%%%%%%%%%%%%%%%%%%%%%%%%
%

As we have discussed above, the $g$-mode frequencies of newly born SQSs are dependent on the masses of quarks. In \Fig{fig:quarkmass} we show the constituent quark masses of the three flavours as functions of the baryon number density, in unit of the saturation baryon number density $\rho_{0}^{}$, for the charge-neutral quark matter at zero temperature and neutrino abundance. We find that with the increase of the density, the dynamical chiral symmetry is restored gradually, and the constituent quark masses of $u$ and $d$ quarks decrease from $350\,\mathrm{MeV}$ to $50\,\mathrm{MeV}$ when the density increases from $0.1\,\rho_{0}^{}$ to $3\,\rho_{0}^{}$. Since the current mass of the strange quark is larger, its constituent mass decreases relatively milder in comparison with the light quarks. Fig.~\ref{fig:EAB} depicts the energy per baryon number $E/A$ and the effective bag constant $B_{\textrm{eff}}^{1/4}$ defined in \Eq{eq:Beff} versus the density for the charge-neutral quark matter at zero temperature.
%
%It is evident that dependence of the $B_{eff}^{1/4}$ on the density obtained here is consistent with %that given in Ref.~\cite{Buballa:2003qv}.
%
One can observe easily from the figure that when the charge-neutral quark matter is in equilibrium, where $E/A$ has a minimum value, $E/A$ is about $1100\,\mathrm{MeV}$, larger than the energy per nucleon in Iron nuclei about $930\,\mathrm{MeV}$. As we will see in the following, the large constituent mass of the $s$ quark leads to its very low abundance in the charge-neutral quark matter. In turn, the Fermi energy is enhanced and the $E/A$ increases.
Furthermore, we find that the effective bag constant $B_{\textrm{eff}}^{1/4}$ increases from about $40\,\mathrm{MeV}$ to $170\,\mathrm{MeV}$ as the density increases from $0.1\,\rho_{0}^{}$ to $3\,\rho_{0}^{}$.  Such a baryon number density dependence of the $B_{\textrm{eff}}^{1/4}$ is qualitatively consistent with that given in Ref.~\cite{Buballa:2003qv}.

It is left to employ the NJL model to construct the configuration of the SQS, for more discussions about the charge-neutral quark matter, see, {\it e.g.}, Ref.~\cite{Buballa:1998pr}. In \Fig{fig:Yimass} we present the abundances of the three flavor quarks and their corresponding constituent quark masses as functions of the stellar radial coordinate for a newly born SQS. One can see that $u$ and $d$ quarks in the NJL model are still relativistic, and their constituent quark masses range from about $30\,\mathrm{MeV}$ to $85\,\mathrm{MeV}$ inside the star. The strange quark in the NJL model, however, is nonrelativistic, since its constituent mass is much larger than those of light quarks, as seen in the right panel of \Fig{fig:Yimass}. It is the large constituent quark mass, which results in very low abundance of strange quarks inside the SQS. Therefore, all the constituents of a SQS, described by the NJL model, are relativistic as well, which are similar with the case in the MIT model. Thus, it is expected that the $g$-mode eigenfrequencies of a SQS, with its EOS described by the NJL model, are also smaller than those of a newly born NS.

\begin{table}[htb]
\begin{center}
\caption{Eigenfrequencies (Hz) of quadrupole ($l=2$) oscillations of $g$-mode for newly born SQSs, with the quark matter described by the NJL model, at three instants of time (ms) after the core bounces.}
\label{tab:NJLgmode}
\begin{tabular}{LMMM}
\hline \hline                    \vspace{0.1cm}
Radial order of $g$-mode  & $t\!=\!100$&$t\!=\!200$&$t\!=\!300$ \\\hline
 $n=1$ & 100.2& 115.4& 107.4    \\
 $n=2$ & 60.1 & 57.0 & 51.8    \\
 $n=3$ & 42.9 & 40.6 & 40.2    \\
 $n=4$ & 31.7 & 31.4 & 29.6    \\
\hline \hline
\end{tabular}
\end{center}
\end{table}

In \tab{tab:NJLgmode} we list some of our obtained eigenfrequencies for the $g$-mode oscillations of newly born SQSs in the NJL model. In the same way, we choose three representative instants of time after the core bounce. We find that the $g$-mode eigenfrequencies of newly born SQSs with $l=2$, $n=1$ in the NJL model are $100.2\,\mathrm{Hz}$, $115.4\,\mathrm{Hz}$, $107.4\,\mathrm{Hz}$ at the three instants, respectively. These values are a bit larger than those of the relevant modes in the MIT bag model as shown in \tab{tab:gmode}, because of the finite constituent masses of light quarks in the NJL model, but they are still much smaller than those of newly born NSs.

\section{\label{sec:sum}Conclusions}

In this work we have studied the nonradial oscillations of newly born NSs and SQSs. The relativistic nuclear field theory with hyperon degrees of freedom is employed to describe the equation of state for the stellar matter in NSs, while both the MIT bag model and the Nambu--Jona-Lasinio model are adopted to construct the configurations of the SQSs. We find that there are no hyperons in the newly born NSs, and it is also found that the $g$-mode eigenfrequencies of newly born SQSs are much lower than those of NSs, no matter which model we choose to describe the newly born SQSs, which implies that the conclusion is model independent. Note, however, that there are some differences between the strange quark matter described by the bag model and that by the NJL model. In the MIT bag model, all the three flavour quarks are relativistic, and the maximal particle mass is the current mass of strange quarks, about $150\,\mathrm{MeV}$, while in the NJL model, quark masses are generated (or dressed) through the DCSB, and the $u$ and $d$ quarks are still relativistic, but the strange quark has large constituent mass. Because of the large constituent mass of the $s$ quark, its abundance in the charge-neutral SQSs is very low, therefore, all the constituents of SQSs are still relativistic, which results in low $g$-mode eigenfrequencies of newly born SQSs.

Furthermore, we have also investigated other modes of nonradial oscillations of newly born NSs and SQSs, such as the $f$- and $p$-modes. We find that eigenfrequencies of the $f$- and $p$-modes are larger than those of $g$-mode, and for the $f$- and $p$-modes, eigenfrequencies of the newly born SQSs are larger than those of the newly born NSs.

In the light of the first direct observation of gravitational waves~\cite{Abbott:2016blz}, and the $g$-mode oscillations of the type II supernovae core serving as potential, efficient sources of gravitational waves~\cite{Ott:2006qp}, it is promising to employ the gravitational waves to identify the QCD phase transition in compact stars, {\it i.e.}, high density strong interaction matter.

\begin{acknowledgments}
The work was supported by the National Natural Science Foundation of China under Contracts No. 11435001; the National Key Basic Research Program of China under Contract Nos. G2013CB834400
 and 2015CB856900; the Fundamental Research Funds for the Central Universities under Contract No. DUT16RC(3)093.
\end{acknowledgments}

%\nocite{*}

%\bibliography{refspec}% Produces the bibliography via BibTeX.
%
%\bibliography{ms}% Produces the bibliography via BibTeX.
%

\end{document}